\documentclass[pra, twocolumn, floatfix]{revtex4}
\usepackage{graphicx}
\usepackage{epstopdf}
\usepackage{amsmath, amsfonts, amssymb, bm,color}
\begin{document}
\title{Performance of the collective three-level quantum thermal engine}
\author{Mihai A. \surname{Macovei}}
\email{mihai.macovei@ifa.md}
\affiliation{Institute of Applied Physics, Academiei str. 5, MD-2028 Chi\c{s}in\u{a}u, Moldova}
\date{\today}
\begin{abstract}
We investigate the performance of a microscopic quantum heat engine consisting of $V-$ or $\Lambda-$type emitters interacting collectively or 
independently when being in contact with environmental thermal reservoirs. Though the efficiency of a Carnot's cycle is always higher than those 
associated with these setups, we have found that the performance of the cooperative $\Lambda-$type heat engine may be larger than that of the 
$V-$type under similar conditions. Cooperativity among the emitters plays an important role for the $\Lambda-$type setup, significantly improving 
its performance, while is less relevant for a $V-$type thermal engine. This is because the population inversion on the working atomic transition as 
well as its off-diagonal elements behave differently for these two atomic ensembles.
\end{abstract}
\maketitle

\section{Introduction}
Quantum heat engines, converting thermal energy into mechanical work, have attracted considerably attention since already many years \cite{thmas,rev}, 
also, in context of laws of quantum thermodynamics applied to small systems \cite{kos1,kos2,nor,varr,wack}. Based on those laws, it was demonstrated 
that the efficiencies of few-level thermal quantum engines are limited by the Carnot efficiency \cite{thmas,rev,kos1,kos2,nor,varr,wack,alicki}, though, 
one may go beyond the Carnot limit via squeezed thermal reservoirs \cite{sqth}. A thorough analysis of a three-level system as amplifiers or attenuators 
was given in \cite{ampl}, while the electromagnetically induced transparency may be used to construct a quantum heat engines \cite{harris}. The quantum 
statistics of a single-atom heat engine was investigated too, in Ref.~\cite{ag_sc}. Substantial work was performed with respect to an experimental realization 
of a quantum thermal engine \cite{exp1,exp2,exp3,exp4,exp5}. Particularly, it was demonstrated that quantum effects are responsible for enhancing the 
output power of a quantum microscopic heat engine compared to that of any classical one using the same resources \cite{exp4,exp5}. Earlier, it was 
emphasized the relevance of quantum effects in extracting work from a single thermal bath \cite{scully1,scully2}. 

The quantum behaviors of a microscopic thermal engine may change considerably if collective phenomena among its elements are taken into account \cite{colw,cols}. 
From this point of view, the role of entanglement in a small self-contained quantum refrigerator was investigated in \cite{pop}, whereas the performances of 
quantum heat engines can be enhanced via collective interactions among many few-level emitters, used as working substance \cite{must,uzdin,epl,kur,otto}. 
Furthermore, the output work may scale quadratically with the number of elements constituting the cooperative thermal engine \cite{must,uzdin}. Collective 
effects greatly enhance the charging power of quantum batteries \cite{qbatt} while the quantum thermometry, that is, the precision of the temperature 
estimation, improves for larger spin ensembles too \cite{lspin}. An ensemble of indistinguishable quantum machines can give rise as well to a genuine 
quantum enhancement of the collective thermodynamic performance \cite{prlp}. Moreover, a quantum Otto cycle in which the medium, an interacting 
ultracold gas, is driven between a superfluid and an insulating phase can outperform similar single particle cycles \cite{bush}.

\begin{figure}[t]
\includegraphics[width = 8cm]{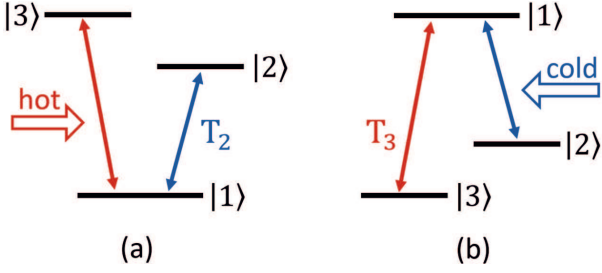}
\caption{\label{fig-1} 
The heat engine setup for a collection of $N$ three-level (a) $V-$ and (b) $\Lambda-$type emitters. The hot bath acts on the 
$|3\rangle \leftrightarrow |1\rangle$, while the cold one on the $|2\rangle \leftrightarrow |1\rangle$ atomic transition, 
respectively. A weak coherent electromagnetic field is applied on the $|3\rangle \leftrightarrow |2\rangle$ transition, converting 
the incoherent thermal energy into a mechanical work.}
\end{figure}

Thus, motivated by the recent substantial progress towards this issue, we investigate here the quantum performance of a microscopic thermal engine 
composed generally from $N$ three-level $V-$ or $\Lambda-$type emitters. More precisely, the working subsystem may consist of single or multiple 
atoms interacting independently or collectively, in the Dicke's sense \cite{dicke,gsa_b,gr_har,gaox,fic_sw,martin}, via the surrounding thermal reservoirs. 
Particularly, the hot bath acts on the $|3\rangle \leftrightarrow |1\rangle$ transition, while the cold one on the $|2\rangle \leftrightarrow |1\rangle$ atomic 
transition, respectively, as it is depicted in Fig.~(\ref{fig-1}), in analogy with the two {\it heat reservoirs} forming the part of a macroscopic classical 
motor operating between two thermal baths. To close the cycle, a weak and coherent electromagnetic field is applied on the $|3\rangle \leftrightarrow |2\rangle$ 
transition, converting the incoherent thermal energy of the heat reservoirs to an output work in an inverted populated atomic medium. Therefore, we are 
interested in the performance of this process in a collective heat engine. Also, we have assumed that the external applied coherent field modifies insignificantly 
the sample's steady-state achieved due to the environmental thermal reservoirs alone. This fact allows us to find the steady-state solutions of the 
corresponding master equations describing the collective three-level samples and, actually, in this way one can investigate the population inversion on the 
working transition and the performance of such a cooperative quantum thermal engine. We have found that generally the performance of a heat engine 
formed from an independent $V-$type atomic ensemble is larger than that when cooperativity among the emitters would be relevant. On the other hand, 
the corresponding performance for collectively interacting $\Lambda-$type emitters is always larger than that for independent atoms. Furthermore, in 
similar conditions, the quantum performance for a $\Lambda-$type ensemble is larger than the one for a $V-$type atomic sample. However, their 
efficiencies are smaller than the efficiency of the Carnot cycle.

The article is organized as follows. In Sec. II we describe the analytical approach and the collective population dynamics of the system of interest, while 
in Sec. III we analyze the corresponding quantum performance of the cooperative three-level microscopic heat engine. The summary is given in Sec. IV.

\section{Collective population dynamics of the microscopic three-level thermal engine \label{s2}}
The master equation describing a collection of $N$ three-level $V-$type ensemble, see Fig.~\ref{fig-1}(a) and Appendix \ref{appA}, interacting with a weak external 
coherent electromagnetic field on the transition $|3\rangle \leftrightarrow |2\rangle$, with the hot thermal bath on the $|3\rangle \leftrightarrow |1\rangle$ transition, 
while the cold one is acting on the $|2\rangle \leftrightarrow |1\rangle$ atomic transition, respectively, in the Born-Markov approximations 
\cite{gsa_b,gr_har,gaox,fic_sw,martin}, is:
\begin{eqnarray}
\frac{d}{dt}\rho(t) &+&  i\Omega[S_{32}+S_{23},\rho] = \nonumber \\
&-&\sum_{\alpha \in \{2,3\}}\frac{\gamma_{\alpha}}{2}(1+\bar n_{\alpha})\bigl\{[S_{\alpha 1},S_{1 \alpha}\rho] + H.c.\bigr \} \nonumber \\
&-& \sum_{\alpha \in \{2,3\}}\frac{\gamma_{\alpha}}{2}\bar n_{\alpha}\bigl\{[S_{1 \alpha},S_{\alpha 1}\rho] + H.c.\bigr\}.
\label{MeqT}
\end{eqnarray}
Here the collective operators $S_{\alpha\beta}=\sum^{N}_{j=1}S^{(j)}_{\alpha\beta}$, with $\{\alpha,\beta \in 1,2,3 \}$ and $S^{(j)}_{\alpha\beta}$=
$|\alpha\rangle_{j}{}_{j}\langle \beta|$, satisfy the commutation relations $[S_{\alpha\beta},S_{\beta'\alpha'}]$=$\delta_{\beta\beta'}S_{\alpha\alpha'}$-
$\delta_{\alpha\alpha'}S_{\beta'\beta}$. The single-atom spontaneous decay rate on transition $|\alpha\rangle \to |1\rangle$ is given by $\gamma_{\alpha}$, 
$\{\alpha \in 2,3\}$, whereas the corresponding mean thermal photon number due to the environmental thermal reservoirs, at temperature $T_{\alpha}$, is 
given by the following expression: $\bar n_{\alpha}=[\exp{\frac{\hbar\omega_{\alpha 1}}{k_{B}T_{\alpha}}}-1]^{-1}$, with $k_{B}$ being the Bolzmann's 
constant and $\omega_{\alpha\beta}=\omega_{\alpha}-\omega_{\beta}$. $\Omega$, considered real, is the corresponding Rabi frequency. 

Since the external coherent electromagnetic field applied on the working transition $|3\rangle \leftrightarrow |2\rangle$ is considered weak, that is 
$\Omega \ll \{N\gamma_{2},N\gamma_{3}\}$, the final steady state is determined mainly by the heat reservoirs. Therefore, the steady-state solution 
of the master equation (\ref{MeqT}), when $\Omega=0$, is given by the following expression:
\begin{eqnarray}
\rho_{s}=Z^{-1}e^{-\xi_{2}S_{22}}e^{-\xi_{3}S_{33}},
\label{ros}
\end{eqnarray}
where $Z$ is determined by the requirement $Tr\{\rho_{s}\}=1$. The substitution of the steady-state solution (\ref{ros}) in the Eq.~(\ref{MeqT}) results in
\begin{eqnarray}
\xi_{\alpha} = \ln\biggl(\frac{1+\bar n_{\alpha}}{\bar n_{\alpha}}\biggr) \equiv -\ln(\eta_{\alpha}),
\end{eqnarray}
where $\eta_{\alpha}=\bar n_{\alpha}/(1+\bar n_{\alpha}) < 1$, $\{ \alpha \in 2,3\}$. 
Notice that the master equation describing a collection of $\Lambda-$type 
three-level emitters, see Fig.~\ref{fig-1}(b), can be obtained from Eq.~(\ref{MeqT}) by swapping the two indices of each transition operator having $\alpha$ or 
$\beta$ as one of the indices, e.g., $S_{\alpha 1} \leftrightarrow S_{1 \alpha}$. The corresponding steady-state solution is given by Eq.~(\ref{ros}) with, however, 
$\xi_{\alpha} = \ln(\eta_{\alpha})$, $\{\alpha \in 2,3\}$ \cite{mmc2}. 

The expectation values of any diagonal statistical moments are obtained by introducing the coherent atomic states $|N, n,m\rangle$ corresponding to the su(3) 
algebra of the operators $S_{\alpha\beta}$ \cite{mmc1,mmc2}. The state $|N, n,m\rangle$ denotes a symmetric collective state of $N$ particles with $n$ atoms 
in bare state $|1\rangle$, $(m-n)$ in bare state $|2\rangle$ and $(N-m)$ atoms in bare state $|3\rangle$. For a given $N$, the admissible values of $\{n,m\}$ 
are $n=0, 1, 2, \cdots, N$ while $m=n, n+1, \cdots, N$.  Thus, the expectation values of the collective population operators, 
$\langle S_{\alpha\alpha}\rangle = -Z^{-1}({\partial}/{\partial \xi_{\alpha}})Z$, $\{\alpha \in 2,3 \}$, with 
\begin{eqnarray*}
Z=\sum^{N}_{n=0}\sum^{N}_{m=n}e^{-\xi_{2}(m-n)}e^{-\xi_{3}(N-m)},
\end{eqnarray*}
needed in the subsequent discussion of the sample's efficiency for a $V-$type three-level ensemble, can be obtained explicitly as:
\begin{eqnarray}
\langle S_{11}\rangle &=& \frac{1}{\eta_{3}-1} \nonumber \\
&-&\frac{(\eta_{2}-\eta_{3})\bigl(1+N-\eta_{2}(N+2) + \eta^{N+2}_{2}\bigr)}{(\eta_{2}-1)\bigl(\eta_{2}-\eta_{3}+(\eta_{3}-1)\eta^{N+2}_{2}
+(1-\eta_{2})\eta^{N+2}_{3}\bigr)}, \nonumber \\
\langle S_{22}\rangle &=& \frac{\eta_{2}}{\eta_{3}-\eta_{2}} \nonumber \\
&+&\frac{\eta_{2}(\eta_{3}-1)\bigl(1+\eta^{N+1}_{2}(\eta_{2}-2 + N(\eta_{2}-1))\bigr)}{(\eta_{2}-1)\bigl(\eta_{2}-\eta_{3}+(\eta_{3}-1)\eta^{N+2}_{2}
+(1-\eta_{2})\eta^{N+2}_{3}\bigr)}, \nonumber \\
\label{peqV}
\end{eqnarray}
and $\langle S_{11}\rangle + \langle S_{22}\rangle+\langle S_{33}\rangle =N$. From the above expressions follow that if $\eta_{2}=\eta_{3}$ then 
$\langle S_{33}\rangle=\langle S_{22}\rangle$. Importantly for the incoming discussions, 
\begin{eqnarray}
\langle S_{33}\rangle > \langle S_{22}\rangle,~~{\it only~ when } ~~ \eta_{3} >\eta_{2}. \label{igV}
\end{eqnarray}
If $\eta_{3}=0$, 
then $\langle S_{33}\rangle =0$, while 
\begin{eqnarray}
\langle S_{22}\rangle = \frac{\eta_{2}(1-\eta^{N}_{2})-N(1-\eta_{2})\eta^{N+1}_{2}}{(1-\eta_{2})(1-\eta^{N+1}_{2})},
\label{ctwo}
\end{eqnarray}
that is, we have recovered the corresponding result for a collection of two-level atoms, $|2\rangle \leftrightarrow |1\rangle$, in a thermal reservoir \cite{colthtwo}.

The corresponding collective steady-state population dynamics for a $\Lambda-$type three-level ensemble is given as follows:
\begin{eqnarray}
\langle S_{11}\rangle &=&\frac{\eta_{2}\eta_{3}\bigl(\eta^{N}_{3}A_{N}(\eta_{2},\eta_{3})-(1-\eta_{2})^{2}\eta^{N}_{2}\bigr)}
{(1-\eta_{2})(1-\eta_{3})D_{N}(\eta_{2},\eta_{3})}, \nonumber \\
\langle S_{22}\rangle &=& \frac{\eta_{3}\bigl( \eta^{N}_{3}B_{N}(\eta_{2},\eta_{3}) +(1-\eta_{2})^{2}\eta^{N+1}_{2}\bigr)}{(1-\eta_{2})(\eta_{3}-\eta_{2})
D_{N}(\eta_{2},\eta_{3})}, \label{peqL}
\end{eqnarray}
where $\langle S_{33}\rangle = N - \langle S_{22}\rangle-\langle S_{11}\rangle$. 
Here, $A_{N}(\eta_{2},\eta_{3})$=$(1-\eta_{3})^{2}+\eta^{N}_{2}\bigl(2-\eta_{2}+N(1-\eta_{2})(1-\eta_{3})-\eta_{3}\bigr)(\eta_{3}-\eta_{2})$, 
$B_{N}(\eta_{2},\eta_{3})$=$(1-\eta_{3})\bigl((2+N)\eta^{2}_{2}+N\eta_{3}-\eta_{2}(1+N)(1+\eta_{3})\bigr)$ - $(\eta_{2} - \eta_{3})^{2}\eta^{N+1}_{2}$,
and $D_{N}(\eta_{2},\eta_{3})$=$(1-\eta_{3})\eta^{N+1}_{3}+(\eta_{2}-1)\eta^{N+1}_{2}+(\eta_{3}-\eta_{2})(\eta_{2}\eta_{3})^{N+1}$. Some limiting 
cases of the steady-state population dynamics, given by (\ref{peqL}), are as follows: when $\eta_{2}\to 0$ or $\eta_{3}\to 0$ then $\langle S_{11}\rangle \to 0$. 
If $\eta_{2}\to 0$ while $\eta_{3} \not=0$, then $\langle S_{22}\rangle \to N$. Conversely, if $\eta_{3}\to 0$ while $\eta_{2} \not=0$, then $\langle S_{33}\rangle \to N$. 
Again here, if $\eta_{2}=\eta_{3}$ then $\langle S_{22}\rangle=\langle S_{33}\rangle$, whereas 
\begin{eqnarray}
\langle S_{22}\rangle > \langle S_{33}\rangle~~{\it only~ if}~~ \eta_{3} >\eta_{2}. \label{igL}
\end{eqnarray}

The cooperative population dynamics given by the expressions (\ref{peqV},\ref{igV},\ref{peqL},\ref{igL}) will be analyzed, in the next Section, in the context of the 
performance of the microscopic thermal engine composed of individual or $N$ three-level emitters, interacting independently or collectively.

\section{The performance of the collective quantum thermal engine}
The maximal quantum efficiency of the microscopic collective thermal engine for both samples (see Fig.~\ref{fig-1}), described in the previous Section, is given by the 
ratio of the energy emitted as work to the energy absorbed from the hot reservoir \cite{thmas,kos2}, namely,
\begin{eqnarray}
\varepsilon = \frac{|\omega_{32}|}{|\omega_{31}|} \label{rand}.
\end{eqnarray}
This efficiency realizes for the transition cycle $|1\rangle \to |3\rangle \to |2\rangle \to |1\rangle$, for $V-$type ensembles, and 
$|3\rangle \to |1\rangle \to |2\rangle \to |3\rangle$ for $\Lambda-$type emitters, respectively, see Fig.~(\ref{fig-1}). However, in the steady-state, the performance
will be governed by the mean values of the populations in these states as well as the induced coherences. Therefore, the output generated work (power output) will be 
proportional to the polarization induced by the applied weak electromagnetic field on the working transition $|3\rangle \leftrightarrow |2\rangle$. Particularly, for a 
$V-$type ensemble the output work is $P_{a}=i\hbar\Omega\langle[\omega_{31}S_{33}+\omega_{21}S_{22},S_{32}+S_{23}]\rangle$=
$i\hbar\Omega(\omega_{31}-\omega_{21})(\langle S_{32}\rangle - \langle S_{23}\rangle)$, whereas for $\Lambda-$type emitters is $P_{a}=i\hbar\Omega\langle[\omega_{13}S_{11}
+\omega_{23}S_{22},S_{32}+S_{23}]\rangle$=$i\hbar\Omega(\omega_{13}-\omega_{12})(\langle S_{23}\rangle - \langle S_{32}\rangle)$, see e.g. \cite{ampl}. In the following, 
we shall relate it with the population inversion between working levels, $\{|3\rangle, |2\rangle\}$, and investigate the performance of the microscopic three-level heat engine 
for individual atoms as well as for independently or collectively interacting emitters, respectively. 

Thus, for a $V-$type atomic ensemble, we have
\begin{eqnarray}
\frac{d}{dt}\langle S_{23}\rangle &=& i\Omega(\langle S_{33}\rangle - \langle S_{22}\rangle)-\gamma_{sV}\langle S_{23}\rangle/2 \nonumber \\
&-& (\gamma_{2}+\gamma_{3})\langle S_{13} S_{21}\rangle/2, \label{pV}
\end{eqnarray}
where $\gamma_{sV}=\gamma_{2}(1+\bar n_{2})$ + $\gamma_{3}(1+\bar n_{3})$, while for the $\Lambda-$type emitters one has:
\begin{eqnarray}
\frac{d}{dt}\langle S_{32}\rangle &=& i\Omega(\langle S_{22}\rangle - \langle S_{33}\rangle)-\gamma_{s\Lambda}\langle S_{32}\rangle/2 \nonumber \\
&+& (\gamma_{2}+\gamma_{3})\langle S_{12} S_{31}\rangle/2, \label{pL}
\end{eqnarray}
with $\gamma_{s\Lambda}=\gamma_{2}\bar n_{2}$ + $\gamma_{3}\bar n_{3}$. One can observe that the last terms of Eqs.~(\ref{pV},\ref{pL}) account for 
the collective effects among the three-level emitters. For larger atomic ensembles, i.e. $N \gg 1$, one can decouple the collective correlator 
$\langle S_{1\alpha}S_{\beta 1}\rangle$, $\alpha\not=\beta$, as follows \cite{TQ}: $\langle S_{1\alpha}S_{\beta 1}\rangle \approx 
\langle S_{11}\rangle \langle S_{\beta\alpha}\rangle$. The decoupling is valid as long as the fluctuations of the population in the $|1\rangle$ 
state, i.e. $\langle \Delta S_{11}\rangle$=$[\langle S^{2}_{11}\rangle - \langle S_{11}\rangle^{2}]/N^{2}$, which may scale as 
$\sqrt{\langle\Delta S_{11}\rangle} \sim N^{-1/2}$, are negligible. This will allow us to obtain the following steady-state expressions for the off-diagonal elements: 
\begin{eqnarray}
\langle S_{23}\rangle =\frac{i\Omega}{\Gamma_{V}}(\langle S_{33}\rangle - \langle S_{22}\rangle), \label{pVV}
\end{eqnarray}
for a $V-$type ensemble, and
\begin{eqnarray}
\langle S_{32}\rangle =\frac{i\Omega}{\Gamma_{\Lambda}}(\langle S_{22}\rangle - \langle S_{33}\rangle), \label{pLL}
\end{eqnarray}
for $\Lambda-$type emitters, respectively. Here, 
\begin{eqnarray}
\Gamma_{V}=(\gamma_{sV}+\gamma_{c})/2,~{\rm while}~\Gamma_{\Lambda}=(\gamma_{s\Lambda}-\gamma_{c})/2, \label{gms}
\end{eqnarray}
with $\gamma_{c}=(\gamma_{2}+\gamma_{3})\langle S_{11}\rangle$ describing the collective contribution. A relevant aspect here is that both the population differences 
as well as $\Gamma_{V,\Lambda}$ depend on the number of emitters when they interact collectively. For single- or many-independent emitters, $\gamma_{c}=0$. Therefore, 
in order to clarify the role played by the cooperativity among the three-level atoms with respect to the output generated work of a many-particle quantum heat engine or its 
efficiency, in the following we investigate the corresponding steady-state population quantum dynamics entering in Exps.~(\ref{pVV},\ref{pLL}). 

For the sake of comparison, firstly, we shall discuss the efficiency of the quantum three-level engine for a single-emitter or an independent atomic ensemble \cite{thmas} 
and, then, for collectively interacting emitters. As we have mentioned in the previous Section, the population quantum dynamics in the steady-state is mainly due to the 
environmental thermal reservoirs, because of the weakness of the applied external electromagnetic field, i.e. $\Omega \ll \{N\gamma_{2},N\gamma_{3}\}$. For a single 
$V-$type atomic system, from Eqs.~(\ref{peqV}) when $N=1$, one obtains:
\begin{eqnarray}
\langle S_{33}\rangle = \frac{\eta_{3}}{1+\eta_{2}+\eta_{3}}, ~{\rm and}~ \langle S_{22}\rangle = \frac{\eta_{2}}{1+\eta_{2}+\eta_{3}}, \label{rrV}
\end{eqnarray}
while their ratio is, respectively,
\begin{eqnarray}
\frac{\langle S_{33}\rangle}{\langle S_{22}\rangle} = \frac{\eta_{3}}{\eta_{2}}. \label{rV}
\end{eqnarray}
Since $\eta_{\alpha}=\exp{[-\hbar \omega_{\alpha 1}/(k_{B}T_{\alpha})]}$, $\{\alpha \in 3,2\}$, one then has that:
\begin{eqnarray}
\frac{\langle S_{33}\rangle}{\langle S_{22}\rangle} = \exp{\bigl[\frac{\hbar\omega_{32}}{k_{B}T_{2}}(\varepsilon_{C}/\varepsilon-1)\bigr]}, \label{rVC}
\end{eqnarray}
where
\begin{eqnarray}
\varepsilon_{C}=1-\frac{T_{2}}{T_{3}}, \label{randC}
\end{eqnarray}
is the Carnot efficiency. Population inversion, i.e. $\langle S_{33}\rangle > \langle S_{22}\rangle$, occurs only if $\eta_{3} > \eta_{2}$ meaning that $\omega_{21}/T_{2} > \omega_{31}/T_{3}$ or, equivalently (see the population ratio (\ref{rVC})),
\begin{eqnarray}
\varepsilon < \varepsilon_{C}.
\end{eqnarray}
\begin{figure}[t]
\includegraphics[width = 4.29cm]{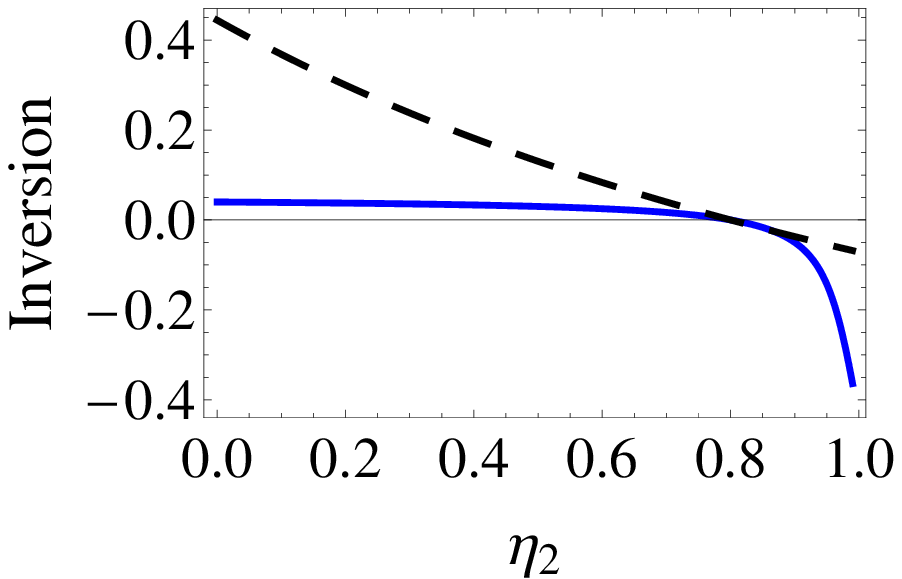}
\hspace{-0.4cm}
\includegraphics[width = 4.35cm]{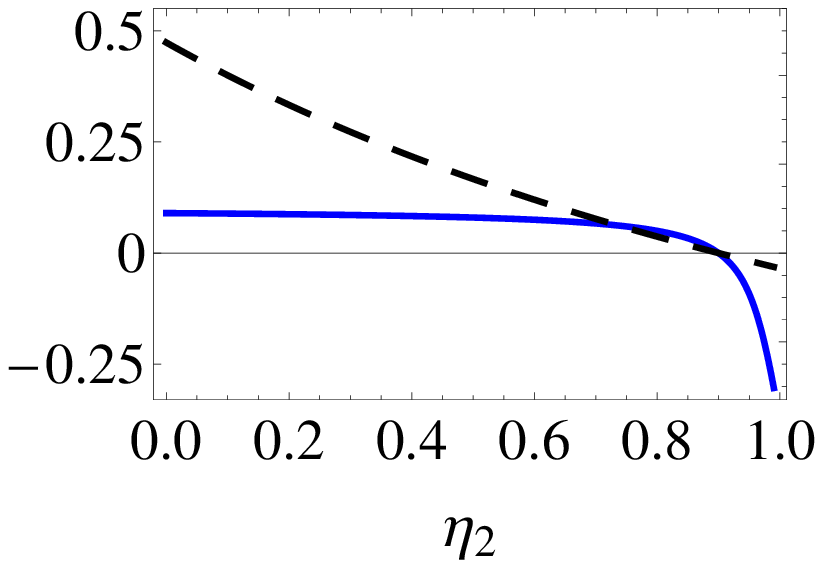}
\begin{picture}(0,0)
\put(-140,67){(a)}
\put(-22,67){(b)}
\end{picture}
\caption{\label{fig-2} 
The scaled steady-state inversion operator $(\langle S_{33}\rangle - \langle S_{22}\rangle)/N$, for $V-$type atomic ensembles, as a function of $\eta_{2}$ 
when (a) $\eta_{3}=0.8$ and (b) $\eta_{3}=0.9$, respectively. The dashed curves describe the situation of a singe-atom or independent atomic ensembles, 
whereas the solid ones depict the case of collectively interacting emitters with $N=100$. Notice that $0 \le \eta_{2} \le 0.99$ here.}
\end{figure}
\begin{figure}[b]
\includegraphics[width = 4.29cm]{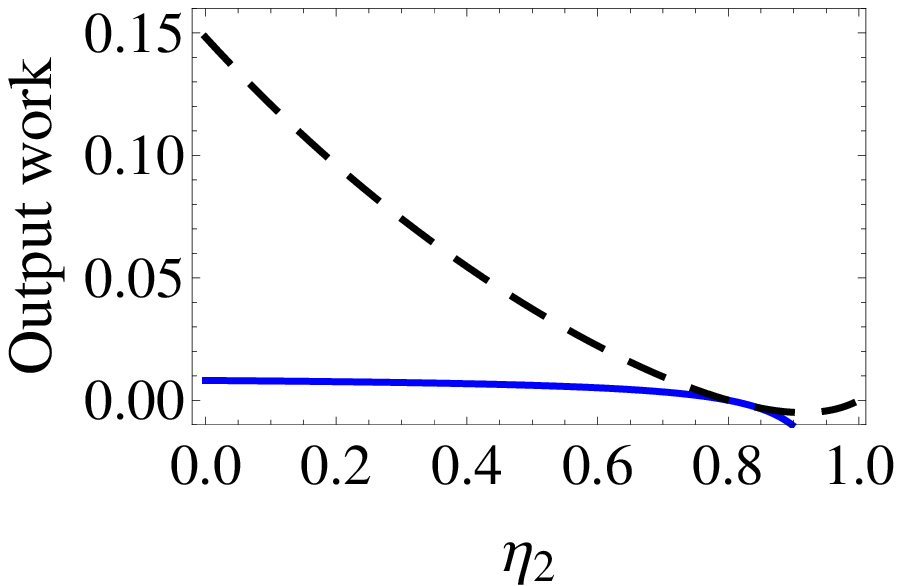}
\hspace{-0.4cm}
\includegraphics[width = 4.35cm]{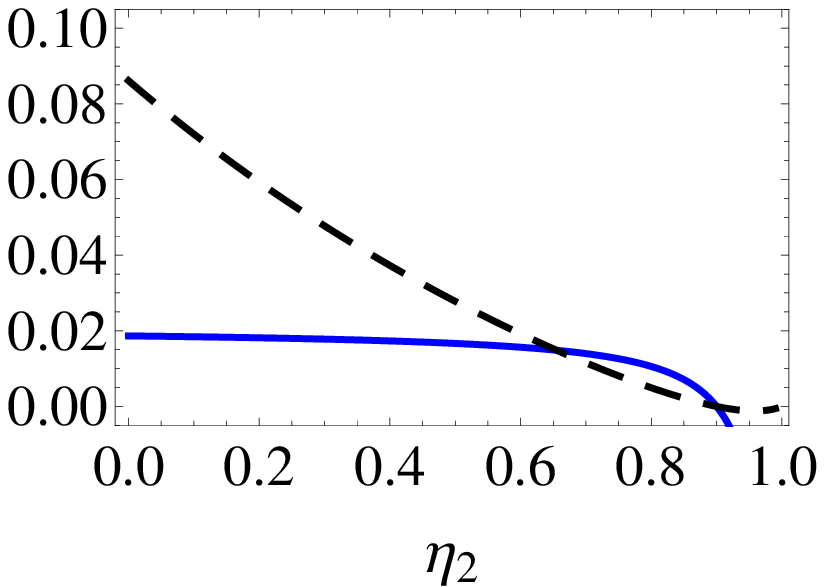}
\begin{picture}(0,0)
\put(-140,68){(a)}
\put(-22,68){(b)}
\end{picture}
\caption{\label{fig-3} 
The scaled output work, proportional to the imaginary part of $\langle S_{23}\rangle/N$ [in units of $\Omega/\gamma$], given by Eq.~(\ref{pVV}) for $V-$type 
atomic ensembles, as a function of $\eta_{2}$ when (a) $\eta_{3}=0.8$ and (b) $\eta_{3}=0.9$, respectively. The dashed curves describe the situation 
of independent atomic ensembles, whereas the solid ones depict the case of collectively interacting emitters. The solid lines were plotted by multiplying with 
a factor equal to $20$. Here, $\gamma_{2}=\gamma_{3}\equiv \gamma$, $N=100$ while $\eta_{2} \le 0.99$.}
\end{figure}

\begin{figure}[t]
\includegraphics[width = 4.29cm]{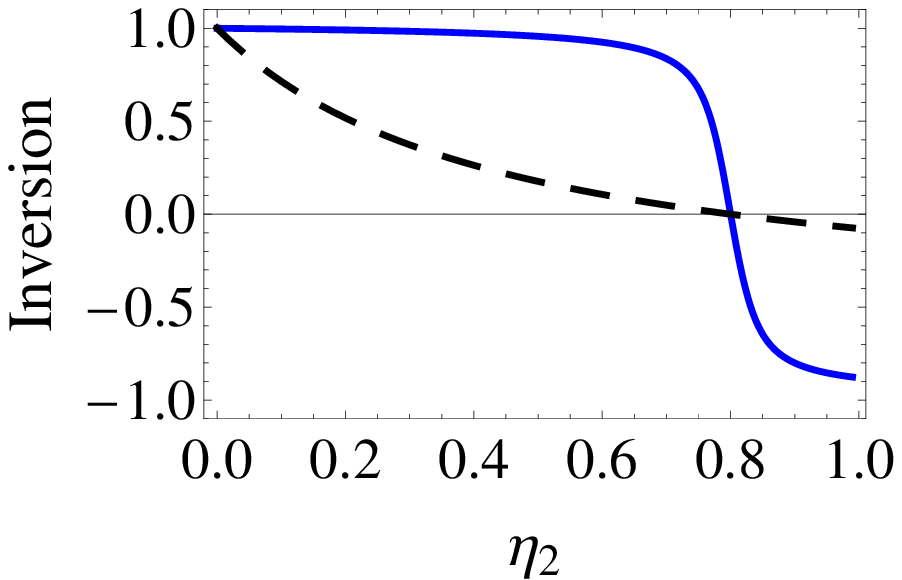}
\hspace{-0.4cm}
\includegraphics[width = 4.35cm]{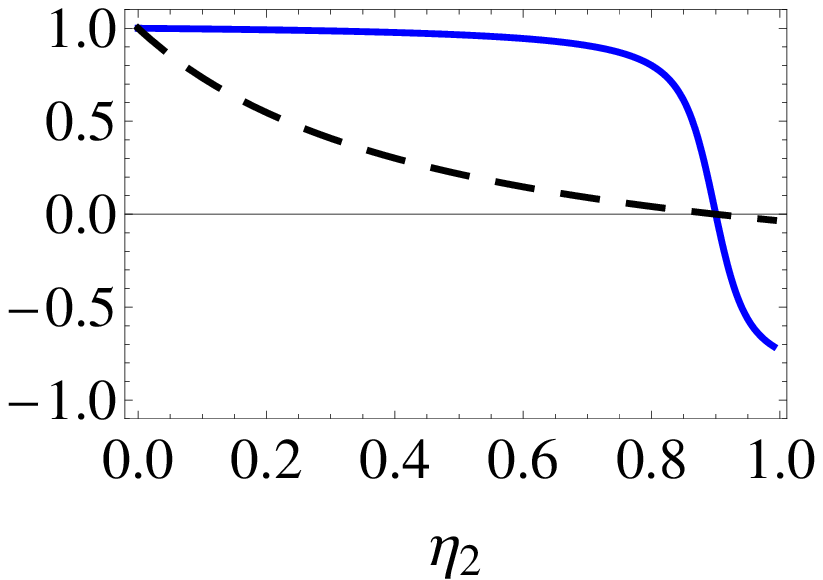}
\begin{picture}(0,0)
\put(-140,67){(a)}
\put(-20,67){(b)}
\end{picture}
\caption{\label{fig-4} 
The scaled steady-state inversion operator $(\langle S_{22}\rangle - \langle S_{33}\rangle)/N$, for $\Lambda-$type atomic ensembles, as 
a function of $\eta_{2}$ when (a) $\eta_{3}=0.8$ and (b) $\eta_{3}=0.9$, respectively. Other parameters are as in Fig~(\ref{fig-2}).}
\end{figure}

Correspondingly, for a single $\Lambda-$type atom, see Fig.~\ref{fig-1}(b), from Eqs.~(\ref{peqL}) when $N=1$, we have \cite{mmc2}:
\begin{eqnarray}
\langle S_{22}\rangle = \frac{\eta_{3}}{\eta_{2}+\eta_{3}+\eta_{2}\eta_{3}}, ~~{\rm and}~~ 
\langle S_{33}\rangle = \frac{\eta_{2}}{\eta_{2}+\eta_{3}+\eta_{2}\eta_{3}}, \nonumber \\
\label{rrL}
\end{eqnarray}
with, however,
\begin{eqnarray}
\frac{\langle S_{22}\rangle}{\langle S_{33}\rangle} = \frac{\eta_{3}}{\eta_{2}}. \label{rL}
\end{eqnarray}
or \cite{thmas}
\begin{eqnarray}
\frac{\langle S_{22}\rangle}{\langle S_{33}\rangle} = \exp{\bigl[\frac{\hbar\omega_{23}}{k_{B}T_{2}}(\varepsilon_{C}/\varepsilon-1)\bigr]}, \label{rLC}
\end{eqnarray}
Here, the population inversion, i.e. $\langle S_{22}\rangle > \langle S_{33}\rangle$, happens only if $\eta_{3} > \eta_{2}$ or $\omega_{12}/T_{2} > \omega_{13}/T_{3}$ 
which, again, implies that $\varepsilon < \varepsilon_{C}$, see the ratio (\ref{rLC}). Notice that for an independent $V-$ or $\Lambda-$type atomic ensemble one should 
multiply the expressions (\ref{rrV}) and (\ref{rrL}) by $N$, with $\sum_{\alpha} \langle S_{\alpha\alpha}\rangle=N$, $\{\alpha \in 1,2,3\}$. Thus, concluding this part, 
the efficiency of such a $V-$ or $\Lambda-$type heat engine consisting of individual atoms or an ensemble of independent emitters will be always smaller than that given 
by the Carnot cycle, i.e., $\varepsilon_{ind} < \varepsilon_{C}$. Furthermore, the output work of the heat engine for $\Lambda-$type single or independent emitters is 
generally higher than the corresponding work for $V-$type emitters, in similar conditions (follow the incoming discussions). This is because the population inversion for 
$\Lambda$ atoms is larger than for $V$ emitters, compare the dashed curves of Fig.~\ref{fig-2}(a) and Fig.~\ref{fig-4}(a) as well as of Fig.~\ref{fig-2}(b) and 
Fig.~\ref{fig-4}(b).
\begin{figure}[b]
\includegraphics[width = 4.29cm]{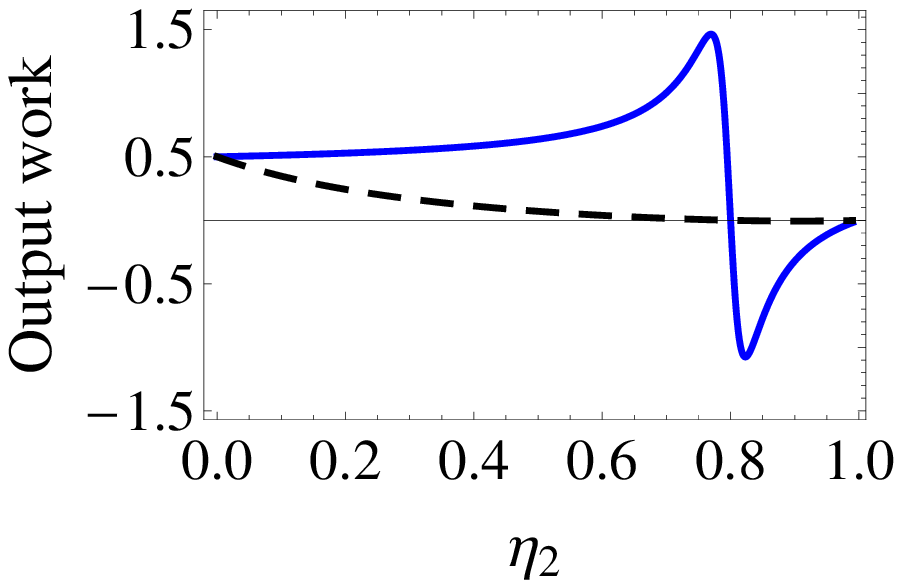}
\hspace{-0.4cm}
\includegraphics[width = 4.35cm]{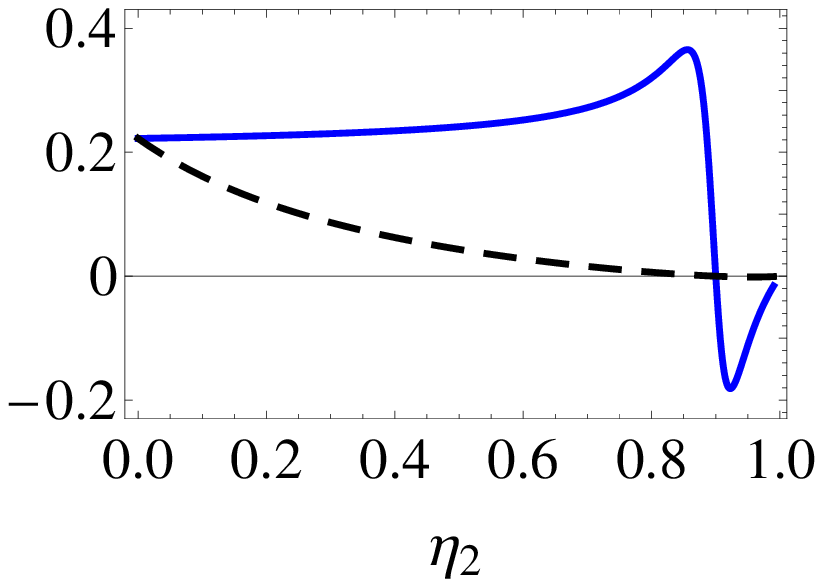}
\begin{picture}(0,0)
\put(-140,68){(a)}
\put(-95,68){(b)}
\end{picture}
\caption{\label{fig-5} 
The scaled output work, proportional to the imaginary part of $\langle S_{32}\rangle/N$ [in units of $\Omega/\gamma$], given by Eq.~(\ref{pLL}) 
for $\Lambda-$type atomic ensembles, as a function of $\eta_{2}$ when (a) $\eta_{3}=0.8$ and (b) $\eta_{3}=0.9$, respectively. The dashed 
lines describe the situation of independent atomic ensembles, whereas the solid ones depict the case of collectively interacting emitters. Here, 
$\gamma_{2}=\gamma_{3}\equiv \gamma$, $N=100$ while $\eta_{2} \le 0.99$.}
\end{figure}

In what follows, we shall focus on a collectively interacting $V-$ or $\Lambda-$type three-level ensemble. According to Eqs.~(\ref{peqV},\ref{peqL}),
the population dynamics of a collection of $N$ three-level emitters depends on $\{\eta_{2},\eta_{3}\}$ in a sophisticated way, but not on their ratio. 
Therefore, simple expressions similar to those given by (\ref{rVC},\ref{rLC}) would not be possible for collectively interacting emitters. On the other 
side, certainly, $\langle S_{33}\rangle > \langle S_{22}\rangle$, for collectively interacting emitters in $V-$type ensembles, while 
$\langle S_{22}\rangle > \langle S_{33}\rangle$, for collectively interacting atoms in $\Lambda-$type samples, {\it only if} $\eta_{3} > \eta_{2}$ as it 
was emphasized in the previous Section. Since the ratio $\eta_{3}/\eta_{2}$ is given by Exps.~(\ref{rVC},\ref{rLC}), one can conjecture then that 
the efficiency of the collective three-level thermal engine, $\varepsilon_{col}$, is smaller than the efficiency of the Carnot cycle, i.e., 
$\varepsilon_{col} < \varepsilon_{C}$. 

While the efficiency of the three-level $V-$ or $\Lambda-$type thermal engine is smaller than that of the Carnot's cycle, regardless of the interaction 
nature among the emitters, the performances of these samples differ if atoms are considered independent or collectively interacting. To elucidate this 
issue, in Figure (\ref{fig-2}), for a $V-$type atomic sample, we plot the scaled inversion operator $(\langle S_{33}\rangle - \langle S_{22}\rangle)/N$
as a function of $\eta_{2}$, while $\eta_{3}$ being fixed. The dashed lines depict the case of independent atoms whereas the solid ones are for 
collectively interacting emitters. Depending on the strength of the hot bath, i.e. $\eta_{3}$, the inversion can be larger for independent or 
collectively interacting atoms, compare Fig.~\ref{fig-2}(a) and Fig.~\ref{fig-2}(b). However, for collectively interacting emitters, the output work 
is inversely proportional to the population into the state $|1\rangle$, see Exps.~(\ref{pVV},\ref{gms}). Therefore, in Figure~(\ref{fig-3}), we plot 
the scaled output work, proportional to the imaginary positive part of $\langle S_{23}\rangle/N$ given by Exp.~(\ref{pVV}) for independently or 
collectively interacting $V-$type emitters. By inspecting this figure, one can observe that generally the output work of a $V-$type heat engine is 
larger for an ensemble of independently interacting emitters rather than if would interact collectively, and it linearly depends on the number of 
emitters (note that the collective curves were obtained by multiplying with $20$). Also, in order to focus on the influence of the thermal 
environmental reservoirs only, we have considered that $\gamma_{2}=\gamma_{3}\equiv\gamma$.

On the other hand, Figure (\ref{fig-4}) depicts the steady-state behaviors of the scaled inversion operator corresponding to the $\Lambda-$type 
ensemble, i.e. $(\langle S_{22}\rangle - \langle S_{33}\rangle)/N$, as a function of the strength of the cold bath $\eta_{2}$ and for different 
values of $\eta_{3}$. In contrast to the $V-$type three-level ensembles, for $\Lambda-$type ones the positive inversion is always higher for 
collectively interacting emitters, compare Fig.~(\ref{fig-2}) and Fig.~(\ref{fig-4}) as well as the dashed and solid lines of Fig.~(\ref{fig-4}), 
respectively. Again here, the output work is inversely proportional to the population in the state $|1\rangle$, see Exps.~(\ref{pLL},\ref{gms}). 
Actually, for collectively interacting $\Lambda-$type emitters this state is less populated for $\eta_{2} < \eta_{3} <1$ and $N \gg 1$. Remarkable 
here, $\Gamma_{\Lambda}$ is smaller for collectively interacting ensembles, see Exp.~(\ref{gms}), meaning that we have an increase in the output 
work from this reason as well as because the inversion enhances too, see Exp.~(\ref{pLL}) and Fig.~(\ref{fig-4}). So, Fig.~(\ref{fig-5}) shows 
the scaled output work, proportional to the imaginary and positive part of $\langle S_{32}\rangle/N$ given by Eq.~(\ref{pLL}) for independently 
or collectively interacting $\Lambda-$type atomic ensembles. We observe here a significant enhance of the output work generated for collectively 
interacting emitters, compared to the independent atoms case. Thus, the performance of a thermal engine based on the collectively interacting 
$\Lambda-$type atoms is larger than for a similar ensemble consisting of independently interacting emitters. Moreover, in a $\Lambda-$type 
ensemble one can almost completely transfer the population to the state $|2\rangle$, see Fig.~(\ref{fig-4}) when $\eta_{2} < \eta_{3}$, meaning 
that in similar conditions the performance of a thermal quantum heat engine consisting of $\Lambda-$type three-level emitters would be higher than 
the performance of a thermal engine formed from $V-$type three-level emitters, compare Fig.~(\ref{fig-3}) and Fig.~(\ref{fig-5}), correspondingly.

Thus, generalizing here, a microscopic $\Lambda-$type three-level quantum heat engine may have an advantage over a similar one formed, 
respectively, of an ensemble of $V-$type three-level emitters. Furthermore, the output work of a cooperative $\Lambda-$type thermal quantum 
engine, i.e. its performance, is larger than that of a heat engine consisting from single or independently interacting $\Lambda-$ atoms. Actually, 
it is greater than the one for single, independently or collectively interacting $V-$type three-level emitters, in similar conditions.

\section{Summary}
Summarizing, we have investigated the efficiency and performance of a microscopic three-level quantum heat engine and have elucidated the role 
the collectivity among the emitters plays with respect to this issue. Since in a cooperative $\Lambda-$type ensemble one can more efficiently create 
population inversion on the involved working atomic transition, the energy conversion of the incoherent thermal reservoirs towards the coherently 
applied electromagnetic field probing that transition is highly improved compared to an independent atomic ensemble or to an independently or 
collectively interacting $V-$type emitters of a microscopic heat engine, under identical conditions. Furthermore, the quantum efficiency of a Carnot 
cycle is always better than that characterizing these setups, regardless of the cooperativity among emitters.

\acknowledgments
Useful discussions with Christoph H. Keitel and J\"{o}rg Evers are gratefully acknowledged as well as the financial support by the 
Moldavian National Agency for Research and Development, grant No. 20.80009.5007.07.

\appendix
\section{Master equation for an ensemble of $V-$type three-level emitters \label{appA}}
The Master equation describing an arbitrary collection of $N$ three-level $V-$type emitters interacting with their environmental reservoirs as well as 
with a coherent electromagnetic field is, see Fig.~\ref{fig-1}(a):
\begin{widetext}
\begin{eqnarray}
\frac{d}{dt}\rho(t) &+&  i\sum^{N}_{j=1}\Omega_{j}\left[S^{(j)}_{32}e^{i\vec k_{L}\cdot \vec r_{j}} + S^{(j)}_{23}e^{-i\vec k_{L}\cdot \vec r_{j}},\rho\right] 
+ i\sum^{N}_{j\not=l=1}\sum_{\alpha \in \{2,3\}}\frac{\gamma_{\alpha}\Omega^{(\alpha)}_{jl}}{2}\left[S^{(j)}_{\alpha 1}S^{(l)}_{1\alpha},\rho\right] \nonumber \\
&=& -\sum^{N}_{j,l=1}\sum_{\alpha \in \{2,3\}}\frac{\gamma_{\alpha}\chi^{(\alpha)}_{jl}}{2}\biggl\{(1+\bar n_{\alpha})
\left[S^{(j)}_{\alpha 1},S^{(l)}_{1 \alpha}\rho\right]  + \bar n_{\alpha}\left[S^{(j)}_{1 \alpha},S^{(l)}_{\alpha 1}\rho\right] \biggr \} + H.c..
\label{MeqA}
\end{eqnarray}
\end{widetext}
For dipole-allowed transitions one has that $\chi^{(\alpha)}_{jl}=\sin{(2\pi r_{jl}/\lambda_{\alpha 1})}/(2\pi r_{jl}/\lambda_{\alpha 1})$ and 
$\Omega^{(\alpha)}_{jl}=\cos{(2\pi r_{jl}/\lambda_{\alpha 1})}/(2\pi r_{jl}/\lambda_{\alpha 1})$, $\alpha \in \{2,3\}$, where we have averaged over all dipole 
orientations, whereas $r_{jl}=|\vec r_{j}-\vec r_{l}|$ are the inter-particle intervals between the $j$th and the $l$th emitters, respectively \cite{gaox,fic_sw,martin}. 
$\lambda_{\alpha 1}$ is the wavelength of the photon emitted on the $|\alpha \rangle \to |1\rangle$ transition, respectively. Further, 
$S^{(j)}_{\alpha\beta}=|\alpha\rangle_{jj}\langle\beta|$, with $\{\alpha,\beta\} \in \{1,2,3\}$, represents the population of the state $|\alpha\rangle$ 
in the $j$-th atom, if $\alpha=\beta$, or the transition operator from $|\beta\rangle$ to $|\alpha\rangle$ of the $j$-th atom when $\alpha\neq\beta$. 
The atomic operators obey the commutation relations 
$\left[S^{(j)}_{\alpha\beta},S^{(l)}_{\beta'\alpha'}\right]=\delta_{jl}\left(\delta_{\beta\beta'}S^{(j)}_{\alpha\alpha'}- \delta_{\alpha\alpha'}S^{(j)}_{\beta'\beta} \right)$.
$\vec k_{L}$ is the wave-vector of the external applied coherent field. The decay rate on the transition $|3\rangle \to |2\rangle$ is significantly smaller than on the other 
involved transitions and, therefore, is not taken into account here.

The coherent evolution of the examined system is described by the second and third terms of the left-side part of the master equation (\ref{MeqA}), 
where the last one describes the dipole-dipole interaction among the emitters. The damping effects are characterized by the right-side part of the 
master equation (\ref{MeqA}). Other parameters are given in Section \ref{s2}. Both the dipole-dipole interactions and the collective damping effects are 
non-negligible if $r_{jl}/\lambda_{\alpha 1} \lesssim 1$, $\alpha \in \{2,3\}$, i.e., when the inter-particle separation are smaller or of the order of the photon 
emission wavelength on the corresponding transition. In this case one can reduce Eq.~(\ref{MeqA}) to the master equation (\ref{MeqT}), describing 
collectively interacting emitters, if one scales the corresponding decay rates $\gamma_{\alpha}\chi^{(\alpha)}_{jl} \to \gamma_{\alpha}\mu_{\alpha} \equiv \gamma_{\alpha}$, 
where $\mu_{\alpha}$, $\alpha \in \{2,3\}$, is a geometric factor, see e.g. \cite{gr_har}. Thus,
\begin{eqnarray}
\frac{d}{dt}\rho(t) &+&  i\Omega[S_{32}+S_{23},\rho] + i\sum_{\alpha \in \{2,3\}}\Omega^{(\alpha)}_{dd}\left[S_{\alpha 1}S_{1\alpha},\rho\right]= \nonumber \\
&-&\sum_{\alpha \in \{2,3\}}\frac{\gamma_{\alpha}}{2}(1+\bar n_{\alpha})\bigl\{[S_{\alpha 1},S_{1 \alpha}\rho] + H.c.\bigr \} \nonumber \\
&-& \sum_{\alpha \in \{2,3\}}\frac{\gamma_{\alpha}}{2}\bar n_{\alpha}\bigl\{[S_{1 \alpha},S_{\alpha 1}\rho] + H.c.\bigr\}.
\label{MeqTA}
\end{eqnarray}
The dipole-dipole interaction term, proportional to $\Omega^{(\alpha)}_{dd}$, was not included in the master equation (\ref{MeqT}) 
since we anticipated that it will not influence the established steady state, i.e. the steady-state solution (\ref{ros}) commutes with 
$S_{\alpha 1}S_{1\alpha}$, $\alpha \in \{2,3\}$. This also means that $\Omega^{(\alpha)}_{dd} < N\gamma_{\alpha}$, $\alpha \in \{2,3\}$, 
that is, the dipole-dipole caused shift among multiple multi-particle states should be smaller than the collective decay rates. As well, in the coherent pumping term 
the exponent factor $e^{\pm i\vec k_{L}\cdot \vec r_{j}} \to 1$ when the wavelength $\lambda_{L}$ of the external coherent pumping source is bigger
than the ensemble's size, i.e., the emitters are in an equivalent position with respect to the coherent driving. Finally, if $j=l=1$, one obtains from Eq.~(\ref{MeqA})
the corresponding master equation for a single $V-$type atom. The master equation for independent emitters, i.e. if $r_{jl}/\lambda_{\alpha 1} \gg 1$,
can be obtained from (\ref{MeqA}) when $j=l$ and $\chi^{(\alpha)}_{jj} \to 1$, while  $\{ \Omega^{(\alpha)}_{jl}, \chi^{(\alpha)}_{jl} \to 0\}$,
$\alpha \in \{2,3\}$. 

In the same way can be analyzed the respective master equation describing $\Lambda-$type emitters, see Fig.~\ref{fig-1}(b).


\begin{thebibliography}{55}
\bibitem{thmas} H. E. D. Scovil and E. O. Schulz-DuBois, Three-Level Masers as Heat Engines, Phys. Rev. Lett. {\bf 2}, 262 (1959).
\bibitem{rev} G. Benenti, G. Casati, K. Saito, and R. S. Whitney, Fundamental aspects of steady-state conversion of heat to work at 
the nanoscale, Phys. Rep. {\bf 694}, 1 (2017).
\bibitem{kos1} E. Geva and R. Kosloff, Three-level quantum amplifier as a heat engine: A study in finite-time thermodynamics,
Phys. Rev. E {\bf 49}, 3903 (1994).
\bibitem{kos2} E. Geva and R. Kosloff, The quantum heat engine and heat pump: An irreversible thermodynamic analysis of the 
three-level amplifier, J. Chem. Phys. {\bf 104}, 7681 (1996).
\bibitem{nor} H. T. Quan, Yu-xi Liu, C. P. Sun, and F. Nori, Quantum thermodynamic cycles and quantum heat engines,
Phys. Rev. E {\bf 76}, 031105 (2007).
\bibitem{varr} V. Singh, Optimal operation of a three-level quantum heat engine and universal nature of efficiency,
Phys. Rev. Res. {\bf 2}, 043187 (2020). 
\bibitem{wack} A. Kalaee and A. Wacker, Positivity of entropy production for the three-level maser,
Phys. Rev. A {\bf 103}, 012202 (2021).
\bibitem{alicki} R. Alicki, The quantum open system as a model of the heat engine, J. Phys. A: Math. Gen. {\bf 12}, L103 (1979).
\bibitem{sqth} J. Klaers, S. Faelt, A. Imamoglu, and E. Togan, Squeezed Thermal Reservoirs as a Resource for a Nanomechanical 
Engine beyond the Carnot Limit, Phys. Rev. X {\bf 7}, 031044 (2017).
\bibitem{ampl} E. Boukobza and D. J. Tannor, Three-Level Systems as Amplifiers and Attenuators: A Thermodynamic Analysis, 
Phys. Rev. Lett. {\bf 98}, 240601 (2007).
\bibitem{harris} S. E. Harris, Electromagnetically induced transparency and quantum heat engine, Phys. Rev. A {\bf 94}, 053859 (2016).
\bibitem{ag_sc} S.-W. Li, M. B. Kim, G. S. Agarwal, and M. O. Scully, Quantum statistics of a single-atom Scovil–Schulz-DuBois heat engine,
Phys. Rev. A {\bf 96}, 063806 (2017).
\bibitem{exp1} J. Ro{$\beta$}nagel, S. T. Dawkins, K. N. Tolazzi, O. Abah, E. Lutz, F. Schmidt-Kaler, and K. Singer, A single-atom
heat engine, Science {\bf 352}, 325 (2016).
\bibitem{exp2} I. A. Martinez, E. Roldan, L. Dinis, D. Petrov, J. M. R. Parrondo, and R. A. Rica, 
Brownian Carnot engine, Nat. Phys. {\bf 12}, 67 (2016).
\bibitem{exp3} Y. Zou, Y. Jiang, Y. Mei, X. Guo, and Sh. Du, Quantum Heat Engine Using Electromagnetically Induced Transparency,
Phys. Rev. Lett. {\bf 119}, 050602 (2017).
\bibitem{exp4} G. Maslennikov, S. Ding, R. Habl\"{u}tzel, J. Gan, A. Roulet, S. Nimmrichter, J. Dai, V. Scarani, and
D. Matsukevich, Quantum absorption refrigerator with trapped ions, Nat. Commun. {\bf 10}, 202 (2019).
\bibitem{exp5} J. Klatzow, J. N. Becker, P. M. Ledingham, C. Weinzetl, K. T. Kaczmarek, D. J. Saunders, J. Nunn, I. A. Walmsley,
R. Uzdin, and E. Poem, Experimental Demonstration of Quantum Effects in the Operation of Microscopic Heat Engines, 
Phys. Rev. Lett. {\bf 122}, 110601 (2019).
\bibitem{scully1} M. O. Scully, Extracting Work from a Single Thermal Bath Via Quantum Negentropy, 
Phys. Rev. Lett. {\bf 87}, 220601 (2001).
\bibitem{scully2} M. O. Scully, M. S. Zubairy, G. S. Agarwal, and H.Walther, Extracting Work from a Single Heat Bath via
Vanishing Quantum Coherence, Science {\bf 299}, 862 (2003).
\bibitem{colw} B. J. de Cisneros and A. C. Hernandez, Collective Working Regimes for Coupled Heat Engines, 
Phys. Rev. Lett. {\bf 98}, 130602 (2007).
\bibitem{cols} J. Jaramillo, M. Beau, and A. del Campo, Quantum supremacy of many-particle thermal machines,
New J. Phys. {\bf 18}, 075019 (2016).
\bibitem{pop} N. Brunner, M. Huber, N. Linden, S. Popescu, R. Silva, and P. Skrzypczyk, Entanglement enhances cooling in microscopic 
quantum refrigerators, Phys. Rev. E {\bf 89}, 032115 (2014).
\bibitem{must} A. \"{U}. C. Hardal, \"{O}. E. M\"{u}stecaplıoglu, Superradiant quantum heat engine,
Scientific Reports {\bf 5}, 12953 (2015).
\bibitem{uzdin} R. Uzdin, Coherence-Induced Reversibility and Collective Operation of Quantum Heat Machines via Coherence Recycling, 
Phys. Rev. Appl. {\bf 6}, 024004 (2016).
\bibitem{epl} H. Vroylandt, M. Esposito, and G. Verley, Collective effects enhancing power and efficiency,
Europhys. Lett. {\bf 120}, 30009 (2017).
\bibitem{kur} W. Niedenzu and G. Kurizki, Cooperative many-body enhancement of quantum thermal machine
power, New J. Phys. {\bf 20}, 113038 (2018). 
\bibitem{otto} M. Kloc, P. Cejnar, and G. Schaller, Collective performance of a finite-time quantum Otto cycle,
Phys. Rev. E {\bf 100}, 042126 (2019).
\bibitem{qbatt} D. Ferraro, M. Campisi, G. M. Andolina, V. Pellegrini, and M. Polini, High-Power Collective Charging of a Solid-State 
Quantum Battery, Phys. Rev. Lett. {\bf 120}, 117702 (2018).
\bibitem{lspin} C. L.  Latune, I. Sinayskiy, and F. Petruccione, Collective heat capacity for quantum thermometry and
quantum engine enhancements, New J. of Phys. {\bf 22}, 083049 (2020).
\bibitem{prlp} G. Watanabe, B. P. Venkatesh, P. Talkner,  M.-J. Hwang, and A. del Campo, Quantum Statistical Enhancement of the Collective 
Performance of Multiple Bosonic Engines, Phys. Rev. Lett. {\bf 124}, 210603 (2020).
\bibitem{bush} Th. Fogarty and Th. Busch, A many-body heat engine at criticality, Quantum Sci. Technol. {\bf 6}, 015003 (2021).
\bibitem{dicke} R. H. Dicke, Coherence in Spontaneous Radiation Processes, Phys. Rev. {\bf 93}, 99 (1954).
\bibitem{gsa_b} G. S. Agarwal, {\it Quantum Statistical Theories of Spontaneous Emission and their Relation to other Approaches} 
(Springer, Berlin, 1974).
\bibitem{gr_har} M. Gross and S. Haroche, Superradiance: An essay on the theory of collective spontaneous emission,
Physics Reports {\bf 93}, 301 (1982).
\bibitem{gaox} J. Peng and G.-x. Li, {\it Introduction to Modern Quantum Optics} (World Scientific, Singapore, 1998).
\bibitem{fic_sw} Z. Ficek and S. Swain, {\it Quantum Interference and Coherence: Theory and Experiments} (Springer, Berlin, 2005).
\bibitem{martin} M. Kiffner, M. Macovei, J. Evers, and C. H. Keitel, Vacuum induced processes in multilevel atoms, 
Prog. Opt. {\bf 55}, 85 (2010).
\bibitem{mmc2} M. Macovei, Y. Niu, S. Gong, and C. H. Keitel, Correlated atomic population fluctuations via the environmental reservoir,
Jr. of Mod. Opt. {\bf 56}, 704 (2009).
\bibitem{mmc1} M. Macovei, J. Evers, and C. H. Keitel, Quantum correlations of an atomic ensemble via an incoherent bath,
Phys. Rev. A {\bf 72}, 063809 (2005).
\bibitem{colthtwo} S. S. Hassan, G. P. Hildred, R. R. Puri, and R. K. Bullough, Incoherently driven Dicke model, 
J. Phys. B: Atom. Mol. Phys. {\bf 15}, 2635 (1982).
\bibitem{TQ} T. Quang and V. Buzek, Squeezing by nondegenerate four-wave mixing in a system
of three-level atoms: effects of the thermal field, J. Opt. Soc. Am. B {\bf 7}, 1487 (1990).
\end{thebibliography}
\end{document}